\documentclass[runningheads, a4paper]{llncs}
	\usepackage{fancyvrb}
	\usepackage{color}

	\usepackage{graphicx}
	\usepackage{amssymb}
	\usepackage{enumerate}
	\usepackage[noend]{algpseudocode}
	\usepackage{algorithm}
	 \usepackage[colorlinks]{hyperref}
	\usepackage{xspace}

	\algdef{SE}[DOWHILE]{Do}{doWhile}{\algorithmicdo}[1]{\algorithmicwhile\ #1}
	
	\newcommand {\R}   {\mathbb R}
	
	\newcommand {\C}   {\mathbb C}

	\newcommand{\ass}{\leftarrow}
	
	\newcommand{\app}[1]{\widetilde{#1}}
	\newcommand{\conj}[1]{\overline{#1}}
	
	\newcommand{\mIm}[1]{Im(#1)}
	\newcommand{\Tstar}[1]{T_*(#1)}
	\newcommand{\contDisc}[1]{\Delta_{#1}}

	\newcommand{\contDiscBox}[1]{\Delta_{#1}}
	\newcommand{\widthCC}[1]{W_{#1}}
	\newcommand{\widthBox}[1]{W_{#1}}

	\newcommand{\solsIn}[2]{\texttt{Zero}(#1,#2)}
    \newcommand{\nbSolsIn}[2]{\#(#1,#2)}

	\newcommand{\cored}[1]{{\color{red}#1}}
	\newcommand{\coblue}[1]{{\color{blue}#1}}

	\newcommand{\ccluster}{\texttt{Ccluster}\xspace}
	\newcommand{\cclusterR}{\texttt{CclusterR}\xspace}
	\newcommand{\cclusterD}{\texttt{CclusterD}\xspace}
	
	\newcommand{\julia}{\texttt{Julia}\xspace}
	\newcommand{\mpsolve}{\texttt{MPsolve}\xspace}
	
	\newcommand{\Ber}{\mbox{\tt Bern}}	
	\newcommand{\Mig}{\mbox{\tt Mign}}
	\newcommand{\Man}{\mbox{\tt Mand}}
	\newcommand{\Spi}{\mbox{\tt Spir}}	
	
	\newcommand{\ignore}[1]{}
	
\usepackage{breqn}
\newtheorem{Definition}{Definition}
\newtheorem{Remark}[Definition]{Remark}

\newtheorem{Proposition}[Definition]{Proposition}

 \title{Polynomial root clustering and explicit deflation}
 \titlerunning{Polynomial root clustering and explicit deflation}

\author{R\'emi Imbach \inst{1}
		\thanks{
		R\'emi's work is supported by NSF Grants \#~CCF-1563942,
    \#~CCF-1564132 and \#~CCF-1708884.
		}
	\and Victor Y. Pan\inst{2}
		\thanks{
	    Victor's work is supported by NSF Grants
	    \#~CCF-1116736 and \#~CCF-1563942 and by PSC CUNY Award
		698130048.}
	    }
\authorrunning{Imbach-Pan}
\institute{
  Courant Institute of Mathematical Sciences\\
  New York University, USA\\
  Email: \email{remi.imbach@nyu.edu}\\
  \url{https://cims.nyu.edu/~imbach/}
\and
  City University of New York\\
  Email: \email{victor.pan@lehman.cuny.edu}\\
  \url{http://comet.lehman.cuny.edu/vpan/}
}

\graphicspath{{./figures/png/}} 
 
\begin{document}
\maketitle
 
\begin{abstract}
%
    We seek complex roots 
    of a univariate polynomial $P$ with real or complex coefficients.
    We address this problem based on recent algorithms that use subdivision
    and have a nearly optimal complexity. 
    They are particularly efficient when only
    roots in a given Region Of Interest (ROI) are sought.
    In this report we explore explicit deflation of
    $P$ to decrease
    its degree and the arithmetic cost of the subdivision.
\end{abstract}


\section{Introduction}

In this report we consider the problem of finding the complex roots 
of a univariate polynomial $P$ with real or complex coefficients.
To address this problem, methods using simultaneous Newton-like iterations
(\emph{e.g.} Erhlich-Aberth iterations) have demonstrated their superiority,
in practice, over other approaches.
Beside the known fact that the convergence of such iterations to solutions 
is not shown, methods based on this idea are global
in the sense that all the roots are found.

In contrast, recent approaches based on the subdivision of an initial box
(the ROI for Region Of Interest) of the complex plane find only
roots in this ROI, which is relevant in many areas of computational
sciences. These methods have also a proved nearly optimal complexity,
and the implementation described in \cite{ICMSpaper}
have shown that they are a little
more efficient for  the local task of
 computing the roots
in a ROI containing only a small number of roots
(which is important in many computational areas) than the
best algorithms for
 global task of approximation of all roots, based on 
Erhlich-Aberth iterations.
These local methods compute clusters of roots, and are robust even in the case  of multiple roots.
To define the Local Clustering Problem
(LCP), let us introduce some definitions.
For any complex set $\mathcal{S}$, 
$\solsIn{\mathcal{S}}{P}$ stands for the roots of $P$ in $\mathcal{S}$,
and $\nbSolsIn{\mathcal{S}}{P}$ for the number of roots, counted
with multiplicity, of $P$ in $\mathcal{S}$.

We consider square complex boxes and complex discs.
If $\mathcal{S}$ is such a box (resp. disc)
and $\delta$ is a positive real number,
we denote by $\delta\mathcal{S}$ the 
box (resp. disc) with the same center than $\mathcal{S}$
but $\delta$ times its width (resp radius).
A disc $\Delta$ is called an \emph{isolator}
if $\nbSolsIn{\Delta}{P}>0$ and 
it is \emph{natural} if in addition
$\nbSolsIn{\Delta}{P}=\nbSolsIn{3\Delta}{P}$.
A set $\mathcal{R}$ of roots of $P$ for which there exist
a natural isolator $\Delta$ with
$\solsIn{\mathcal{R}}{P}=\solsIn{\Delta}{P}$
is called a \emph{natural cluster}.
The LCP is to compute natural isolators for natural
clusters together with the sum of multiplicities of roots in the
clusters:

\begin{center} \fbox{
		\begin{minipage}{0.9 \textwidth} \noindent
		\textbf{Local Clustering Problem (LCP):}\\ \noindent
		\textbf{Given:} a polynomial $P\in\C[z]$, 
		                a square complex box $B_0\subset\C$, 
		                $\epsilon>0$\\
		\noindent \textbf{Output:} a set of pairs
		$\{(\Delta^1,m^1),\ldots,(\Delta^\ell,m^\ell)\}$ where: \\ \noindent
		\hphantom{\textbf{Output:}} - the $\Delta^j$s are pairwise disjoint
		discs of radius $\leq\epsilon$,\\ \noindent
		\hphantom{\textbf{Output:}} - each $m^j =
		\nbSolsIn{\Delta^j}{P}=\nbSolsIn{3\Delta^j}{P}$ 
		and $m^j>0$\\ \noindent
		\hphantom{\textbf{Output:}} - $\solsIn{B_0}{P} \subseteq
		\bigcup_{j=1}^{\ell} \solsIn{\Delta^j}{P} \subseteq
		\solsIn{2B_0}{P}$.
	\end{minipage}
	} \end{center}

We present here a practical improvement
obtained by 
successively deflating $P$.
Once a set $\mathcal{S}$ of $\nbSolsIn{\mathcal{S}}{P}$
roots of $P$ counted with multiplicities has been found,
one can compute the factor $Q$ of $P$
that has exactly the roots 
$\solsIn{\C}{P}\setminus\solsIn{\mathcal{S}}{P}$
with the same multiplicities,
then compute clusters of roots of $Q$
that has a smaller degree than $P$.

\subsubsection*{Previous works}

Univariate polynomial root finding is a long standing quest 
that is still actual; it is intrinsically linked to polynomial
factorization for which the theoretical record upper bound, which differs from an information lower bound  by at most a polylogarithmic factor in the input size 
 has been achieved in \cite{pan2002univariate}. 
Root-finder supporting such bit
complexity bounds are said nearly optimal.
User's choice,
 however, has been for a while
the package of subroutines \mpsolve (see \cite{bini2000design}and \cite{bini2014solving}), based on simultaneous Newton-like 
(\emph{i.e.} Ehrlich-Aberth iterations).
These iterations converge  to all roots simultaneously with cubic convergence rate, but only locally, that is, near 
the roots, and empirically converge very fast globally, with no formal support known for this empirical behavior.
Furthermore they compute a small number of roots in a ROI
not much faster than all roots.

In contrast, recent approaches based on subdivision 
compute the roots in a fixed ROI at the cost that decrease at least proportionally to the number of roots.
In the case where only the real roots are sought,
subdivision can be mixed with the Descartes rule of signs 
and Newton iterations (\cite{sagraloff2016computing})
to achieve a near optimal complexity.
The implementation described in \cite{Kobel}
demonstrated the practical efficiency of this approach.

In the complex case, a subdivision method with a nearly optimal 
complexity have also been proposed in \cite{2016Becker}.
This method computes natural clusters and is robust
in the case of multiple roots; its implementation (\cite{ICMSpaper})
is a little more efficient than \mpsolve for ROI's
containing only several roots; when all the roots are sought,
\mpsolve remains the user's choice.

A recent study of polynomial deflation can be found in 
\cite{pan2018polynomial}.

\section{Root clustering with explicit deflation}


\subsubsection*{The base root clustering algorithm}

We rely here on a procedure 
\[
clusterPol(Q,\mathcal{D},\epsilon,\mathcal{C}, n)
\]
based on the reduction of a research domain $\mathcal{D}$
taking as input:
\begin{itemize}
 \item a polynomial $Q$ satisfying $\solsIn{\C}{Q}\subseteq\solsIn{\C}{P}$ given as an oracle,
 \item the search domain $\mathcal{D}$,
 \item an $\epsilon>0$,
 \item a list $\mathcal{C}$ of pairwise disjoint $\epsilon$-clusters of roots of $P$
       in $\C\setminus\mathcal{D}$ and
 \item an integer $n$.
\end{itemize}
It finds at most $n$ $\epsilon$-clusters of roots of $Q$ in $\mathcal{D}$
and reduces the search domain.
More precisely,
it returns a list $\mathcal{C}_*$
of $\ell$ pairwise disjoint $\epsilon$-clusters of roots of $Q$
and a domain $\mathcal{D}_*\subset\mathcal{D}$ 
so that:
\begin{enumerate}[$(i)$]
 \item $\solsIn{\mathcal{D}}{Q}\subseteq\solsIn{\mathcal{C}_*\cup\mathcal{D}_*}{Q}$,
 \item $\solsIn{\mathcal{C}_*}{Q}\subseteq\solsIn{\mathcal{D}}{Q}$,
 \item either $\ell=n$, 
       or $\mathcal{D}_*$ is empty,
 \item elements in $\mathcal{C}_*\cup\mathcal{C}$ are pairwise disjoints.
%
\end{enumerate}

Such a procedure can be implemented for instance with an algorithm based on 
box quadri-section,
in which case the search domain will be a queue of boxes
that are leaves in the subdivision tree of $B_0$.
\begin{Remark}
 \label{rem_correct}
 Let $\mathcal{C}_*,\mathcal{D}_*$ be the result of $clusterPol(Q,\mathcal{D},\epsilon,\mathcal{C}, n)$
where $Q$ is such that $\solsIn{\C}{Q}\subseteq\solsIn{\C}{P}$.
If $\mathcal{D}_0$ is such that $\mathcal{D}\subset\mathcal{D}_0$
and $\mathcal{C}$ contains all the roots of $P$ in $\mathcal{D}_0\setminus\mathcal{D}$,
then $\mathcal{C}\cup\mathcal{C}_*$ contains all the roots of $P$ in 
$\mathcal{D}_0\setminus\mathcal{D}_*$; 
if in addition $\mathcal{D}_*$ is empty, $\mathcal{C}\cup\mathcal{C}_*$
is a solution for the LCP for $P$, $\mathcal{D}_0$, $\epsilon$.
\end{Remark}

We also rely on a procedure 
\[
 refine(\mathcal{C}, L)
\]
taking as an input 
a list $\mathcal{C}$ of pairwise disjoints natural $\epsilon$-clusters of roots of $P$ 
and an integer $L>1$,
and returning a list $\mathcal{C}_*$
of pairwise disjoints natural $2^{-L}$-clusters of roots of $P$
so that $\solsIn{\mathcal{C}}{P}=\solsIn{\mathcal{C}_*}{P}$;
$refine(\mathcal{C}, L)$ possibly splits clusters in $\mathcal{C}$.

\subsubsection*{Root clustering with explicit deflation}

We present in Algo.~\ref{algo:ClusDAC} our main procedure for 
computing clusters of roots of $P$ with explicit deflation.
At each re-entrance in the {\bf while} loop in step 2,
$\mathcal{C}$ contains natural $\epsilon$ clusters of roots of $P$
that are in $\mathcal{D}_0\setminus\mathcal{D}$, and all the 
roots of $P$ in $\mathcal{D}_0\setminus\mathcal{D}$ are in 
$\mathcal{C}$.
Let $Q$ be the unique monic polynomial that has exactly the roots
of $P$ that are not in $\mathcal{C}$, with the same multiplicities
as in $P$.
An oracle for $Q$ is obtained in step 3 by specializing
for arguments $P,\mathcal{C}$ the 
procedure $OracleForQ$ defined in Algo.~\ref{algo:Clus2Factor}.
This procedure uses power sums of roots of $P$.
Provided that $OracleForQ$ is correct, the correctness
of Algo.~\ref{algo:ClusDAC} is a direct consequence of 
rem.~\ref{rem_correct}.

\begin{algorithm}
	\begin{algorithmic}[1]
	\caption{$ClusterWithDeflation(P,\mathcal{D}_0,\epsilon,n)$}
	\label{algo:ClusDAC}
	\Require{An oracle for a polynomial $P$, a ROI $\mathcal{D}_0$, $\epsilon>0$, $n\geq1$.} 
	\Ensure{Natural $\epsilon$-clusters of $P$ in $\mathcal{D}_0$.}
	\State $\mathcal{C}, \mathcal{D}\ass clusterPol(P,\mathcal{D}_0,\emptyset,\epsilon, n)$
	\While{$\mathcal{D}$ is not empty}
        \State $Q\ass OracleForQ(P,\mathcal{C},.)$
        \State $\mathcal{C}_*, \mathcal{D}\ass clusterPol(Q,\mathcal{D},\epsilon,\mathcal{C}, n)$
        \State $\mathcal{C}\ass \mathcal{C}\cup\mathcal{C}_*$
    \EndWhile
    \State \Return $\mathcal{C}$
	\end{algorithmic}
\end{algorithm}

\subsubsection*{Power sums of roots}
For a polynomial $P$ and a set $S$ of roots (given with multiplicities) of $P$,
the first $n$ power sums of the roots in $S$
are the $n$-dimensional vector $(a_1,\ldots,a_n)$
where $a_i=\sum\limits_{\alpha\in S} \nbSolsIn{\alpha}{P}\times\alpha^i$.
In the case where $P$ is given by its coefficients, one 
can compute the first $n$ power sums of all its roots 
for $n\leq d_P$ (where $d_P$ is the degree of $P$),
with Newton identities.
Here we will assume the existence of a procedure
\[
 CoeffsToPS(P,n,L)
\]
taking as an input an oracle for a polynomial $P$, 
a precision $L\geq 1$ and an integer $n\geq1$
and returning $L$-bit approximations 
for the first $n$ power sums of all the roots of $P$.

Conversely, given an $n$-dimensional vector $(a_1,\ldots,a_n)$
whose $i$-th component is the $i$-th power sum of $d$ complex numbers
$(\alpha_1,\ldots,\alpha_d)$ with $d\leq n$, one can 
compute the unique monic polynomial $Q$ of degree $d$ having the 
$\alpha_i$'s as its roots.
Then again, one can apply Newton identities.
Here we assume the existence of a procedure
\[
 PSToCoeffs((\app{a_1},\ldots,\app{a_n}),L,d)
\]
taking in input $L$-bit approximations $(\app{a_1},\ldots,\app{a_n})$ for the first $n$ power sums
of $d$ complex numbers $(\alpha_1,\ldots,\alpha_d)$
and returning 
a pair $(\app{Q},L')$ where $\app{Q}$ is  
an $L'$-bit approximation 
for the unique monic polynomial $Q$ of degree $d$
having $(\alpha_1,\ldots,\alpha_d)$ as roots.

%

\subsubsection*{Polynomial deflation with power sums}
Given a set $\mathcal{C}$ of clusters of roots of $P$, we use power sums 
to compute an oracle for the unique monic polynomial $Q$
whose set of roots is exactly 
$\solsIn{\C}{P}\setminus\solsIn{\mathcal{C}}{P}$
with the same multiplicities than in $P$.
First, the degree $d$ of $Q$ is $d_P - \nbSolsIn{\mathcal{C}}{P}$.
Now if $(a_1,\ldots,a_d)$ are the first $d$ power sums
of the roots of $P$ and $(b_1,\ldots,b_d)$
are the first $d$ power sums of the roots of $P$ in $\mathcal{C}$,
then $(a_1-b_1,\ldots,a_d-b_d)$ are the 
$d$ first power sums of the roots of $Q$
and the coefficients for $Q$ can be computed
from these power sums.
The procedure $OracleForQ(P,\mathcal{C}, L)$
described in Algo.~\ref{algo:Clus2Factor}
turns this reasoning into an oracle for $Q$.
The power sums of the roots of $P$ and 
the roots of $P$ in $\mathcal{C}$ are only known
as oracles; one can increase the precision
asked from those oracles until the computed 
polynomial $Q$ has the precision $L$ asked from an input.

In step 8, we suppose that error bounds are computed 
while carrying out the arithmetic operations that return 
the pair $(\app{c_s},L_s)$
meaning that $\app{c_s}$ is an $L_s$-bit approximation of the
result.

%

\begin{algorithm}
	\begin{algorithmic}[1]
	\caption{$OracleForQ(P,\mathcal{C},L)$}
	\label{algo:Clus2Factor}
	\Require{An oracle for a polynomial $P$, a set $\mathcal{C}$ of clusters of roots of $P$, a precision $L\geq 1$.} 
	\Ensure{An $L$-bit approximation for the unique monic polynomial $Q$ 
	        of degree $d_P - \nbSolsIn{\mathcal{C}}{P}$ whose set of roots is exactly $\solsIn{\C}{P}\setminus\solsIn{\mathcal{C}}{P}$ 
            with the same multiplicities as in $P$.}
	\State $d_Q\ass d_P - \nbSolsIn{\mathcal{C}}{P}$
	\State $L_{temp}\ass L$, $L_{res}\ass 0$
	\While{$L_{res}<L$}
        \State $L_{temp}\ass 2L_{temp}$
        \State $\{(\Delta_j,m_j)|1\leq j \leq \ell\}\ass refine(\mathcal{C}, L_{temp})$ 
               \coblue{{\it // the $c(\Delta_j)$'s are $L_{temp}$-bit approx. for the roots of $P$ in $\mathcal{C}$}}
        \State $(\app{a_1},\ldots,\app{a_{d_Q}})\ass CoeffsToPS(P,d_Q,L_{temp})$ 
               \coblue{{\it // $L_{temp}$-bit approx. for the $d_Q$ first PS of all the roots of $P$}}
        \For{$s$ in $1,\ldots,d_Q$}
            \State $(\app{c_s},L_s)\ass \app{a_s} - \sum\limits_{j=1}^{\ell}m_j\times (c(\Delta_j))^{s}$ 
            \coblue{{\it // $L_s$-bit approx. for the $s$-th PS of $Q$, with $L_s<L_{temp}$}}
        \EndFor
        \State $(\app{Q},L_{res})\ass PSToCoeffs((\app{c_1},\ldots,\app{c_{d_Q}}),min_{s} L_s, d_Q)$
	\EndWhile
	\State \Return $\app{Q}$
	\end{algorithmic}
\end{algorithm}

\subsubsection*{Implementation}

We implemented the procedures $ClusterWithDeflation$
and $OracleForQ$
in \julia.
For the procedure $clusterPol$,
we used a modified version of \ccluster, implementing a depth first search
in the subdivision tree.
For the procedure $refine$, we used \ccluster.
We will denote by \cclusterD our prototype implementation of $ClusterWithDeflation$.
We also incorporated improvements described in previous section:
for polynomials in $\R[z]$, \ccluster is \cclusterR.

\subsubsection*{Numerical results}

\begin{figure}
	 \centering
	  \includegraphics[width=6cm]{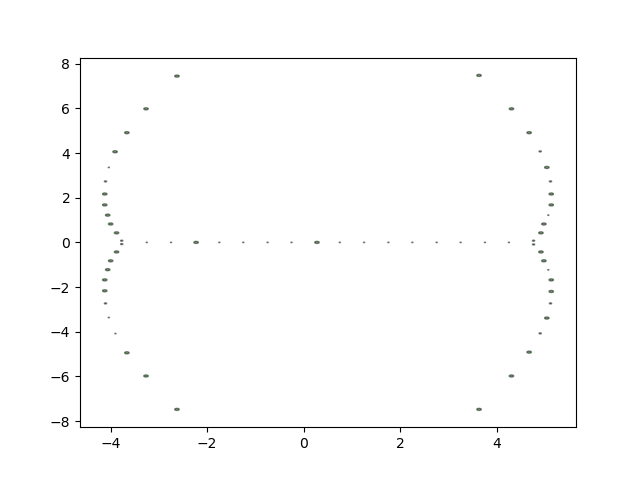}
	    \caption{
	    Clusters of roots, all containing one root, for the Bernoulli polynomial of degree 64.
		  }
	 \label{fig:bernoulli_sols}
	\vspace*{-1em}
	\end{figure}
	
\begin{figure}
	 \begin{minipage}{0.5\linewidth}
	 \centering
	  \includegraphics[width=6cm]{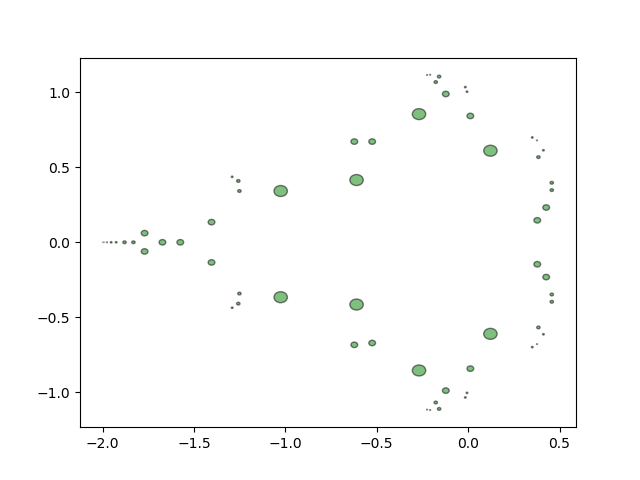}
	 \end{minipage}
	 \begin{minipage}{0.5\linewidth}
	 \centering
	  \includegraphics[width=6cm]{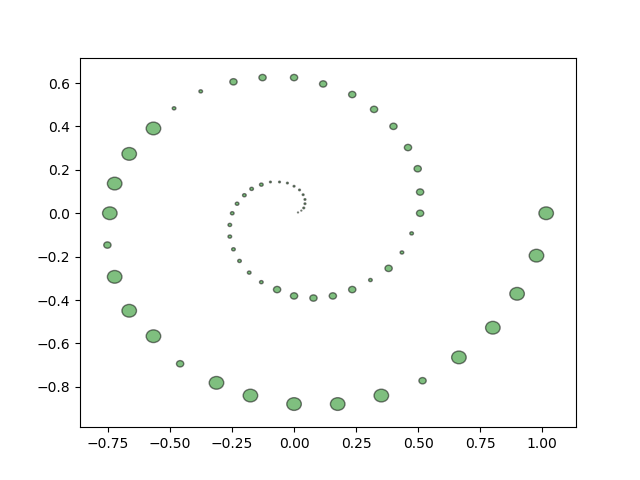} 
	 \end{minipage}
	    \caption{
	    {\bf Left:} 
	    Clusters of roots, all containing one root, for the Mandelbrot polynomial of degree 63.
	    {\bf Right:} Clusters of roots, all containing one root, for the Spiral polynomial of degree 64.
		  }
	 \label{fig:dac}
	\vspace*{-1em}
	\end{figure}

We compare running times of \ccluster and \cclusterD
for finding $\epsilon$-clusters of polynomials in $\R[z]$ and $\C[z]$.
We used three families of polynomials: 
\begin{itemize}
 \item Bernoulli polynomials
	$\Ber_d(z) = \sum_{k=0}^{d} {{d}\choose{k}}b_{d-k}z^k$ where
	$b_i$'s are the Bernoulli numbers,
 \item Mandelbrot polynomials (see \cite{bini2000design}); let $P_0(z)=1$ and
       consider the sequence of polynomials 
       \begin{equation}
       \label{eq:mandel}
       P_k(z) = zP_{k-1}(z)P_{k-1}(z) +1
       \end{equation}
       We define $\Man_d(z)$ as $P_{\lfloor log_2(d+1) \rfloor}(z)$,
       
 \item Spiral Polynomial
		$\Spi_d(z) =
		    \prod_{k=1}^d(z - \dfrac{k}{d}e^{\frac{4k\sqrt{-1} \pi}{n}})$.
\end{itemize}
The roots of Bernoulli polynomial of degree 64 are drawn in 
fig.~\ref{fig:bernoulli_sols}
Mandelbrot and Spiral polynomials of degrees 63 and 64 are drawn 
in fig.~\ref{fig:dac}.

\begin{table}[t]
\begin{center}

\begin{tabular}{l||c||lr|lr|c|c|c|} 
 & \multicolumn{1}{|c||}{\ccluster} 
 & \multicolumn{7}{|c|}{\cclusterD} \\
 & $t_1$ & $t_2$ & $t_1/t_2$ & $t_3$ & $t_3/t_2$ & $t_4$ & $t_5$ & maxprec \\\hline 
Bernoulli  $d=128$ & 5.44 & 4.05 & \cored{1.34} & 0.18 & 0.05 & 0.10 & 0.05 & 424\\\hline 
Bernoulli  $d=256$ & 33.7 & 25.0 & \cored{1.35} & 1.18 & 0.05 & 0.56 & 0.36 & 848\\\hline 
Bernoulli  $d=512$ & 192  & 144  & \cored{1.33} & 10.1 & 0.07 & 5.42 & 2.95 & 1696\\\hline 
\hline 
Mandelbrot $d=127$ & 5.97 & 4.11 & \cored{1.45} & 0.18 & 0.04 & 0.11 & 0.04 & 424\\\hline 
Mandelbrot $d=255$ & 32.3 & 23.3 & \cored{1.38} & 1.47 & 0.06 & 0.89 & 0.36 & 848\\\hline 
Mandelbrot $d=511$ & 212  & 149  & \cored{1.42} & 15.6 & 0.10 & 10.7 & 3.04 & 1696\\\hline 
\hline 
Spiral     $d=128$ & 15.5 & 9.13 & \cored{1.70} & 0.52 & 0.06 & 0.35 & 0.09 & 424\\\hline 
Spiral     $d=256$ & 93.0 & 63.3 & \cored{1.47} & 4.19 & 0.07 & 2.47 & 1.04 & 848\\\hline 
Spiral     $d=512$ & 560  & 423  & \cored{1.32} & 85.1 & 0.20 & 36.1 & 27.7 & 3392\\\hline 
\hline 
\end{tabular}

\end{center}
\caption{Comparison of running times of \ccluster (without deflation) 
         and \cclusterD (with deflations) for polynomials
         with increasing degree $d$.}
\label{table_deflation}
\end{table}

In table~\ref{table_deflation}, 
column $t_1$ and $t_2$ give the sequential times in seconds for 
respectively \ccluster and \cclusterD.
Column $t_3$ gives the cumulative time spent in $OracleForQ$.
Column $t_4$ gives the cumulative time spent in $refine$
and column $t_5$ the cumulative time spent in $PSToCoeffs$.
Column maxprec gives the maximum precision required 
for clusters of roots of $P$
(\emph{i.e.} minus the $log_2$ of the size of the isolating 
disk).
Timings are obtained on a 
Intel(R) Xeon(R) CPU E5-2680 v2 @ 2.80GHz machine with Linux.
For \cclusterD, we choosed $n=\lfloor \frac{d}{8}\rfloor$ that seems to give the best ratios.
For both \ccluster and \cclusterD, 
we choosed $\epsilon=2^{-53}$.
The ROI for Bernoulli polynomials is $[-150,150]+\sqrt{-1}[-150,150]$,
the one for Mandelbrot polynomials is $[-10,10]+\sqrt{-1}[-10,10]$,
and the one for Spiral polynomials is $[-2,2]+\sqrt{-1}[-2,2]$.
In all cases, \ccluster and \cclusterD found all the roots 
in clusters of one root.

\section{Future works}
The present document reports our firt study
of the practical use of deflation approaches
for root finding.
In the future we may experiment less naive 
uses of explicit deflation via power sums computation.
Knowing a disc isolating roots of a factor,
one can recover this factor (without computing the roots)
by computing power sums of those roots, as described in
\cite{pan2018polynomial}, appendix A
(see also \cite{schonhage1982fundamental}).
\bibliographystyle{alpha}
\bibliography{references}

\end{document}